# Absorption and Dispersion: In Search of a Versatile Spectroscopic Technique

**Iskander Gazizov[1], Davide Pinto[1,2], Harald Moser[1], Savda Sam[1,3], Pedro Martín-Mateos[4], Liam O'Faolain[3], and Bernhard Lendl[1,*]**

[1] Institute of Chemical Technologies and Analytics, TU Wien, 1060 Vienna, Austria
[2] Institute for Quantum Electronics, Department of Physics, ETH Zürich, 8093 Zürich, Switzerland
[3] Centre for Advanced Photonics and Process Analysis, Munster Technological University, Bishopstown, T12 P928 Cork, Ireland
[4] Electronics Technology Department, Universidad Carlos III de Madrid, 28911 Leganés, Madrid, Spain
* Correspondence: bernhard.lendl@tuwien.ac.at

**Abstract**:
This paper investigates three spectroscopic techniques — Tunable Diode Laser Absorption Spectroscopy (TDLAS), Wavelength Modulation Spectroscopy (WMS), and Heterodyne Phase Sensitive Dispersion Spectroscopy (HPSDS) — to evaluate their performance in detecting carbon monoxide (CO) in nitrogen in a 31 m Herriot cell using an Interband Cascade Laser (ICL) emitting at 4559 nm. With the developed spectrometer, we can switch between the techniques with minimal hardware modifications. Calibration curves were established using simple peak-to peak values from the recorded TDLAS, WMS and HSPDS spectra as well as after applying respective spectral models of the sensor responses followed by least square fitting approach. We compare the results when using the different techniques with respect to obtained linearity, minimum detectable value (MDV), standard deviation of the method ($s_{x0}$), coefficient of variation of the method ($V_{x0}$), limit of detection (LOD), and Allan-Werle deviation analysis. The results show significant differences in the performance of the three techniques when establishing calibration curves based on peak-to-peak evaluation of the recorded spectra. However, when establishing model-based calibration curves a more uniform performance of the investigated techniques was found. The results demonstrate that TDLAS is a straightforward and robust technique but has a limited measurement range of up to 65 ppm. WMS offers a linear range of up to 100 ppm if an appropriate spectral model is applied, and a baseline-free operation but requires careful calibration. HPSDS stands out for its wide linear range of up to 100 ppm and immunity to power fluctuations. However, its operation requires a complex mathematical model and radio frequency (RF) components. Allan-Werle deviation analysis demonstrates comparable sensitivities of 0.17 ppb for TDLAS, 0.25 ppb for WMS, and 0.20 ppb for HPSDS. Finally, the minimum detectable value from the calibration curves is 15 ppb for TDLAS, 13 ppb for WMS, and 18 ppb for HPSDS.

# Introduction

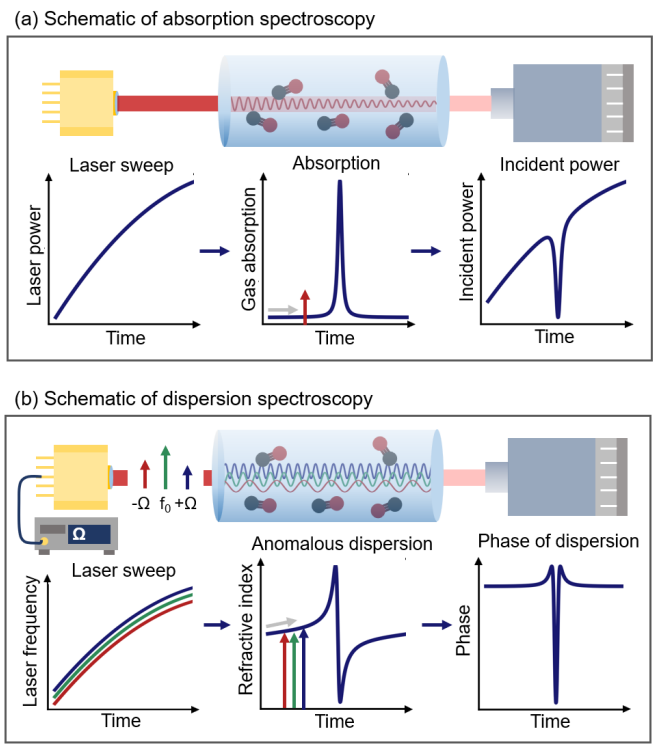

**Fig. 1.** Principle of laser absorption and dispersion spectroscopy techniques: (a) The laser wavelength is tuned across characteristic absorption lines of the target gas, causing a reduction of the measured signal intensity due to absorption, which is detected by a photodiode, and then used to determine the gas concentration. (b) High-frequency ($\Omega$) modulation of the laser causes side-band generation. When these closely frequency-spaced optical tones pass the sample, they are phase-shifted if they experience different refractive indices. Such phase shifts occur when the three tones are swept over an absorption line due to the anomalous dispersion. To measure the phase shift, the tones are mixed on a square law photodetector, where a beat note is produced as a product of the interference.

As noted by the inventor of the laser, Theodore Maiman, "*a laser is a solution seeking a problem*" [1]. Indeed, sixty-four years later, countless problems have been addressed using laser technology. In particular, the application of lasers in the field of gas sensing has allowed instruments to become portable, such as the SAM spectrometer searching methane on Mars [2], and sensitive, like the SCAR-14 radiocarbon dating spectrometer capable of concentration measurements at parts-per-quadrillion (ppq) level [3]. However, with over a dozen widely used laser spectroscopy techniques available, a question arises: which one should be applied to tackle the problems in the future [4,5]?

There are a growing number of ways to implement laser-based gas sensors, including both direct and indirect methods, where some use optical cavities for signal enhancement while others do not [4]. In this work, we focus on direct measurement techniques such as Tunable Diode Laser Absorption Spectroscopy (TDLAS), Wavelength Modulation Spectroscopy (WMS), and Heterodyne Phase Sensitive Dispersion Spectroscopy (HPSDS), which can all be implemented using the same optical setup and hardware components, along with the same data acquisition and processing tools. By concentrating on these fundamental techniques, we believe we can provide a fair and balanced comparison of their performance. To compare the three techniques, we calculated their metrological figures of merit following ISO standards [6,7] and the EURACHEM guidelines [8] which describe the statistical evaluation of calibration functions as used in analytical chemistry.

## Absorption Spectroscopy

Absorption spectroscopy is a well-established method for the non-destructive measurement of gas concentration, temperature [9], and velocity [10]. Absorption-based spectrometers have demonstrated their robustness in different environmental and industrial applications. They are also frequently mounted as sensors on various platforms, including drones [11], meteorological balloons [12], and used for planetary research of the Solar System [2,13–15].

These sensors work by passing narrowband light through an absorbing gas and detecting the remaining light intensity on the detector. Due to the resonance with rotational-vibrational energy levels of the molecules, the light intensity is reduced when the laser is scanned across an absorption line. This reduction is described by the Bouguer-Lambert-Beer law:

$$I(\tilde{v})/I_0(\tilde{v}) = \tau = e^{-\alpha(\tilde{v})L} = e^{-N\sigma(\tilde{v})L} \qquad (1)$$

Where transmittance $\tau$ is the ratio of light intensity before $I_o$ ($\tilde{v}$) and after $I$ ($\tilde{v}$) the gas sample, $\alpha$ (cm$^{-1}$) is the wavenumber-dependent linear absorption coefficient, $L$ (cm) is the length of the light path, $N$ (molecule/cm$^3$) is the number density of the absorbing species derived from the ideal gas law, $\sigma$ (cm$^2$/molecule) is the wavenumber dependent absorption cross section, and $\tilde{v}$ (cm$^{-1}$) is the wavenumber [16].

Some of the most popular absorption techniques are Tunable Diode Laser Absorption Spectroscopy (TDLAS) and Wavelength Modulation Spectroscopy (WMS) [17]. In TDLAS, the laser light sweeps across the absorption feature, and the photodiode detects intensity $I(\tilde{v})$ which is the convolution of the transmission spectrum and laser intensity modulation due to the current ramp, as presented in Fig. 1(a). Extracting the transmission spectrum is challenging and requires accurate estimation of $I_o$ ($\tilde{v}$), also called baseline removal, which can be done by polynomial fitting [18], wavelet transforms [19] or orthogonal polynomials [20,21]. In the early days of TDLAS, a sweep rate of a few tens of Hertz could be achieved, lasers are nowadays swept at the kHz rate to reduce the 1/f noise, and sweeps up to 2 MHz were demonstrated [22]. An advantage of TDLAS is that the calculated absorption spectrum can be directly fitted using spectral line parameters, which results in calibration-free quantification of target gases.

WMS involves applying additional high frequency (kHz-MHz) [23] sinusoidal modulation to the laser. The radiation passes through gas cell, is detected by a square-law detector and the recorded signals are demodulated using a lock-in amplifier (LIA). This process shifts the information to higher frequencies, and by doing so it reduces the 1/f noise, which is strongest at lower frequencies [24]. Another feature of WMS is the ability to lock the laser to the peak of the line which results in faster acquisition times [25]. Generally, the recorded data in WMS can be evaluated by demodulation at the harmonics of the applied modulation frequency. At the second harmonic, baseline offset and slope can be removed efficiently. Extracted WMS harmonics are, however, sensitive to laser power fluctuations. A common solution is to normalise the second harmonic (2*f*) signal with the first harmonic (1*f*) component, known as the WMS-2*f*/1*f* method [24]. It shall be noted that this normalisation procedure is only fully valid in case a completely linearly tuning diode laser is employed in WMS. Extraction and fitting of the obtained spectra are more complicated in WMS in contrast to TDLAS. Nevertheless, WMS is also often described as being calibration-free, however, it requires careful management of the laser's operational parameters due to the added modulation to extract valid absorption spectra as needed for spectral fitting [23,26–29].

Frequency Modulation Spectroscopy (FMS) is similar to WMS but operates at frequencies from 100 MHz to 1 GHz, thus at frequencies significantly higher than WMS, which typically operates below 1 MHz [20].

Such high frequency modulation produces frequency-spaced sidebands of the laser, and the absorption feature is probed by a single isolated sideband [17]. FMS implementation requires additional RF components but allows for measurements of absorption and dispersion derived from frequency and amplitude modulation of the signal caused by the interaction with the spectral feature [17].

## Dispersion Spectroscopy

Absorption and dispersion are both governed by the sample's complex refractive index, with real part $n(\tilde{v})$ providing the dispersion spectrum, and imaginary part $k(\tilde{v})$ being directly proportional to the absorption spectrum [16].

$$\hat{n}(\tilde{v}) = n(\tilde{v}) - ik(\tilde{v}) \qquad (2)$$

Dispersion spectroscopy measures the concentration of an analyte gas by focusing on the wavenumber-dependent refractive index $n(\tilde{v})$ rather than the absorption coefficient $\alpha(\tilde{v})$, which is proportional to the imaginary part of the complex refractive index $k(\tilde{v}) = \alpha(\tilde{v})/(4\pi\tilde{v}_0)$. Here $\tilde{v}_0$ is the mean wavenumber for the spectrum [16].

In gases, each rotational-vibrational absorption line is associated with anomalous dispersion. Specifically, the refractive index is higher on the low-frequency side of the resonance and lower on the high-frequency side. This induces a concentration-dependent phase shift between closely spaced laser lines as they pass the gas sample, as seen in Fig. 1(b) [16]. Conveniently, the dispersion spectrum can be related to the absorption spectrum through the Kramers-Kronig relations [30–32]:

$$n(\tilde{v}) - n(\infty) = \frac{1}{\pi} P \int_{-\infty}^{\infty} \frac{\kappa(\vartheta)}{\vartheta - \tilde{v}} d\vartheta \qquad (3)$$

Here, the refractive index $n(\infty)$ is assumed to be 1, and $\kappa(\tilde{v})$ is the imaginary part of the complex refractive index [16]. The right expression in (3) represents the Hilbert transform $H$, symbol $P$ denotes that the Cauchy principal value is to be taken. To convert any spectrum of absorption coefficient to the refractive index, one could apply the following equation [31]:

$$n(\tilde{v}) - 1 = H[\alpha(\tilde{v})/(4\pi\tilde{v}_0)] \qquad (4)$$

The two recently developed dispersion techniques for gas sensing in IR spectroscopy are Chirped Laser Dispersion Spectroscopy (CLaDS), and Heterodyne Phase Sensitive Dispersion Spectroscopy (HPSDS). In CLaDS, the molecular dispersion signature is deduced from the frequency variation of the heterodyne beat note between two waves originating from a frequency-chirped single laser source [33].

The HPSDS technique does not rely on a chirped laser. It operates by measuring the phase shifts caused by anomalous dispersion near a spectral line [34]. A high-frequency modulation Ω (MHz-GHz) is applied to the laser to generate sidebands, shown in Fig. 1(b) as a three-tone beam. When tuned across the spectral feature of interest, each tone experiences a slightly different refractive index, resulting in phase shifts between the tones. A heterodyne reception is applied to quantify these phase shifts. The three tones interfere on a photodiode, producing an RF beat note at the same modulation frequency Ω. Depending on the implementation, the beatnote is usually downmixed to the lower frequency range, so its phase is demodulated via a lock-in amplifier. The magnitude of the recorded phase shift is proportional to the gas concentration and can be derived from the HPSDS signal as described in [35,36]. The key difference between HPSDS and FMS is that, in FMS the applied modulation is such that only one sideband scans the line profile, whereas in HPSDS, all of the sidebands scan across the line. Also, HPSDS targets the phase shifts between the tones and not the intensity.

To probe the anomalous dispersion both HPSDS and CLaDS require a two- or three-colour beam which is typically created by external modulators. In the absence of modulators for pure intensity modulation in the mid-IR region, lasers are usually modulated directly by means of the driving current [37]. This approach results in mixed intensity modulation (IM) and frequency modulation (FM) [38,39]. Mixed modulation leads to asymmetry in amplitude and phase for the sidebands, creating a nonlinear relationship between the output phase and gas concentration. Adjusting the laser's operating point can suppress the influence of frequency modulation to some extent, as FM-IM ratios vary with modulation frequency and laser bias current [34,36,38].

Key advantages of dispersion techniques that are often mentioned in the literature are the linear response to concentration changes even for strongly absorbing samples, immunity to power fluctuations, and baseline free operation [34,37,40]. In this work, we observe how direct laser modulation introduces unwanted effects in the measurements which reduce the measurement range and introduce a baseline. As will be shown, adjustment of the laser's operating point allows to extend the working range for analyte quantification.

## Carbon Monoxide

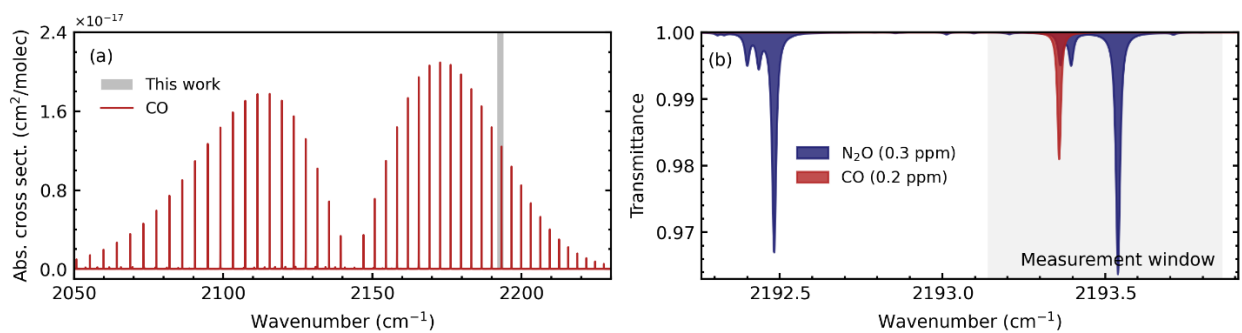

**Fig. 2.** (a) Absorption cross section of the fundamental vibration of carbon monoxide (b). Simulated transmission spectrum of natural air at 0.1 atm over an optical path of 31 m. Only CO and $N_2O$ lines are present in the measurement region highlighted in grey. The overlap between CO and $N_2O$ lines does not affect the measurements because calibrations were carried out on a pure CO+$N_2$ mixture.

In this paper, we compare the performance of the chosen spectroscopic techniques on the example of measuring carbon monoxide (CO) in nitrogen. Carbon monoxide is a colourless and odourless gas mainly produced by incomplete combustion of fossil fuels and natural gas; it is also produced by human respiration [41]. The average CO concentration in cities is less than 1 part-per-million (ppm) [42]. CO is also an excellent tracer of atmospheric transport, with its mixing ratio rapidly decreasing from the troposphere, where its lifetime is only a few months, to the stratosphere, where it can remain for several months at lower concentrations of tens of ppb [43]. Historically, atmospheric CO measurements required sampling and analysis using gas chromatography. Nowadays, non-dispersive infrared (NDIR) and laser-based methods have become popular [44]. In Table 1 we present a comparison between different laser-based techniques targeting CO, with the best 1σ precision of 40 ppt, which is often retrieved by measuring a blank gas sample. As shown in Table 1, this work achieved satisfactory results in terms of precision, and excellent performance in dynamic range.

For our study, we selected the operating range over the R(13) transition of the fundamental vibration band near 2193.36 cm$^{-1}$. The expected spectrum, shown in Fig. 2, was simulated using the HITRAN2020 database [45].

Table 1. Comparison of laser-based spectrometers for CO sensing. TDLAS is Tunable Diode Laser Absorption Spectroscopy. OA-ICOS is Off-Axis Integrated Cavity Output Spectroscopy. CRDS is a Cavity Ring-Down Spectroscopy. QEPAS is Quartz-Enhanced Photoacoustic Spectroscopy. PTS is Photothermal Spectroscopy.

|   | Spectrometer | Spectroscopy technique | Pathlength | **Precision 1σ[1]** | Measurement range |
|---|---|---|---|---|---|
| 1 | Review of C. Zellweger et. al. [44] | TDLAS | 76 m | 0.04 ppb (60 s) | 1.5 ppm |
| 2 | Los Gatos Research Inc. (LGR), USA, model LGR-23d [46] | OA-ICOS | Not defined, but in km scale | 0.05 ppb (180 s) | 10 ppm |
| 3 | Valéry Catoire et. al. [47] | TDLAS | 134 m | 0.06 ppb (150 s) | 405 ppb |
| 4 | Ligang Shao et. al. [48] | TDLAS | 76 m | 0.092 ppb (200 s) | 2 ppm |
| **5** | **This work** | **TDLAS** | **32 m** | **0.14 ppb (60 s)** | **100 ppm** |
| 6 | Picarro, USA, model G2401 [49] | CRDS | Not defined, but in km scale | 0.4 ppb (300 s) | 5 ppm |
| 7 | Silvia Viciani et. al. [43] | TDLAS | 36 m | 0.8 ppb (6 s) | 200 ppb |
| 8 | Davide Pinto et. al. [50] | QEPAS | Not defined, but in mm scale | 6 ppb (100 s) | 100 ppm |
| 9 | Davide Pinto et. al. [50] | PTS | Not defined, but in mm scale | 15 ppb (100 s) | 100 ppm |

## Comparisons of the Investigated Techniques

A comprehensive introduction to laser-based gas sensing is provided in [4,5]. Numerous comparisons of individual techniques exist. For example, simulations of WMS and TDLAS systems demonstrate that both techniques can yield similar results if the TDLAS repetition rate is sufficiently high [51]. In situations when baseline determination is challenging, both TDLAS and WMS offer comparable results [52]. A proposed combination of TDLAS and WMS detection scheme shows WMS outperforming TDLAS at lower concentrations but being nonlinear at higher concentrations [53]. Several studies show that while HPSDS is linear over a wider concentration range compared to WMS, it often has a worse limit of detection [34,36,54]. In another study, it was discussed how HPSDS was limited by the noise from the electronic jitter, and WMS was limited by the normalisation of the signal to the laser power [40].

In this paper, we compare all three techniques: classic WMS and TDLAS with the less explored HPSDS. The comparison focuses on linearity analysis targeting both low and high concentrations where transmittance saturation occurs. From the calibration function, we calculate important analytical figures of merit such as linear range, standard deviation of the method, and variation coefficient. We also calculate the minimum detectable value (MDV) from the calibration [55]. Furthermore, the limit of detection (LOD) is calculated as three times the standard deviation of the blank/low concentration as advised by the EURACHEM Guide [8]. We also investigate the long-term stability of the different techniques using the Allan-Werle deviation analysis [56]. By this method the longest averaging time still leading to a reduction in noise is obtained thus providing a further estimate of achievable sensitivities upon prolonged signal averaging.

By experimenting with the same setup, we aim to present limitations and benefits of each technique in an objective way, providing clarity for researchers in the field of what to expect when adopting one or the other technique. The comparison between simple peak-to-peak evaluation of recorded spectra and more complex non-linear fitting algorithms highlights the impact of data analysis methods on the analytical figures of merit of the established calibration curves. Furthermore, we present new insights into the HPSDS technique, including what we believe to be the first demonstration of direct RF modulation on the Interband Cascade Laser for three-tone generation. Additionally, we propose a novel detection scheme, which allows us to choose between a wide linear range or high sensitivity by controlling the laser's FM-IM properties. This switching technique overcomes a limitation in the detection sensitivity of the HPSDS method, as noted in previous studies [34,36,54].

---

[1] There is no standardized way of calculating precision followed by all authors, thus the comparison is approximate.

# Methods

## Layout of the employed laser-based gas sensor

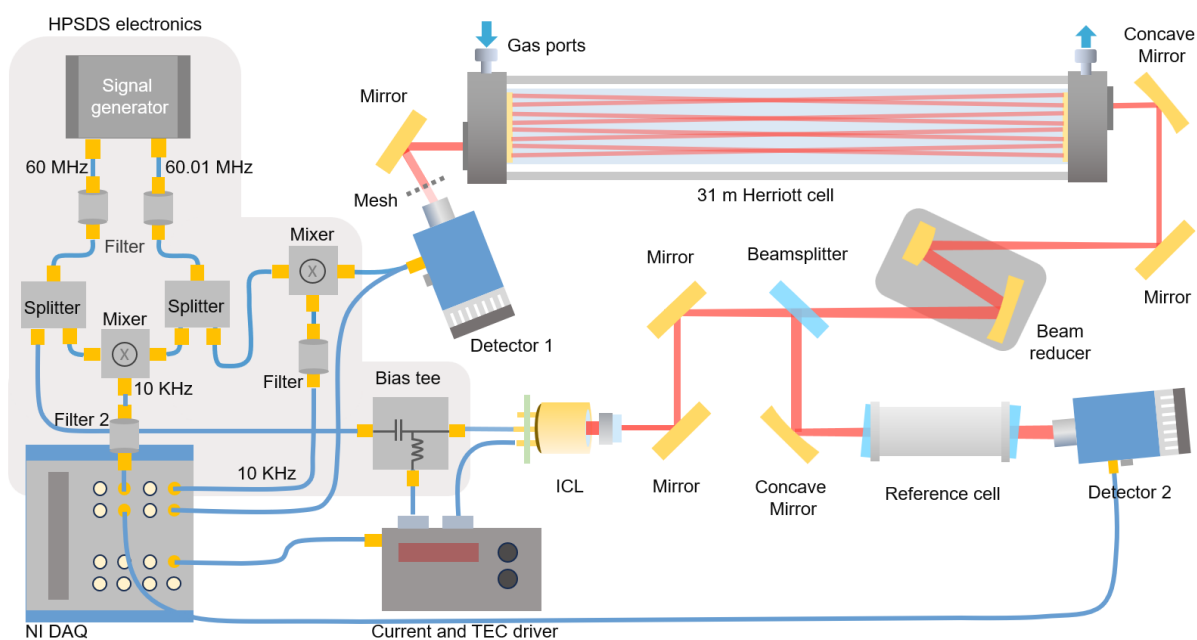

**Fig. 3.** Diagram of the instrument used to test absorption and dispersion techniques. RF components responsible for modulation and demodulation in HPSDS mode are highlighted in grey. In WMS and TDLAS modes, the signal from the photodiode was directly connected to the NI DAQ card, and the laser did not experience RF modulation. ICL is Interband Cascade Laser.

As can be seen from Fig. 3, the design of the instrument represents a classical laser-based spectrometer. For gas sensing, we used a commercially available Herriot cell (HC30L/M-M02, Thorlabs, United States) with a 31.227 m optical path. This cell was chosen for its reasonable path length and straightforward alignment. A 500 mm concave mirror and a 2x beam reducer (BE02R/M, Thorlabs) ensured the required beam diameter for the cell. The light source was an Interband Cascade Laser (ICL) from Nanoplus, Germany, with a 4 mm collimating lens, and emitting at 2193.5 cm$^{-1}$. The laser driver (TLD001, Thorlabs) was operated as a current source, and not a current sink, to simplify RF signal mixing in a bias tee for the HPSDS mode. Laser temperature was controlled using a TEC-1091 from Meerstetter, Switzerland.

The light, after exiting the cell, was detected by an MCT detector. For TDLAS and WMS, we used a low-bandwidth detector (PIP-DC-20M-F-M4, VIGO Photonics, Poland). A fine metal mesh was installed in front of the detector to reduce incident power and prevent nonlinear response. For HPSDS, we used a 1 GHz bandwidth VIGO detector (FIP-1k-1G-F-M8-D) to demodulate the RF beatnote. In this case no mesh was required for signal attenuation.

HPSDS modulation and demodulation involved additional RF components from Mini Circuits (United States). The ~ 60 MHz RF signal from a SynthHD v2 generator (Windfreak Technologies, United States) was mixed with a current sweep in a low inductance bias tee (ZFBT-352-FT+), this created a three-tone signal in the laser output. As these three-tone signals were swept across the spectral feature of interest, each of them experienced a different phase shift. To extract phase information, the tones were received by the photodiode, which produced the beatnote at the modulation frequency of the laser. Next, to demodulate the phase of the RF beatnote with a software lock-in-amplifier (LIA), we first had to lower the beatnote frequency from the MHz range down to 10 kHz, due to the limited bandwidth of LIA. Downmixing was done with a ZFM-4-S+ mixer and an additional reference generator set 10 kHz apart from the main generator. The same mixing was done with a reference generator to produce a 10 kHz reference frequency for the lock-in-amplifier. For signal splitting, which was required for further mixing, we used a ZESC-2-11+ splitter. Additional SLP-300+ filters were installed to suppress unwanted harmonics coming from the RF generator.

For data acquisition, we chose the NI USB-6366 card (National Instruments, United States), as it provided a sample rate of 2 Msps with a resolution of 16-bit, necessary for TDLAS signal averaging and software lock-in-amplifier implementation. The instrument was controlled by LabVIEW software, with final data processing performed in Python. The entire instrument was enclosed to minimize air and temperature fluctuations.

For all measurements, we used two $CO+N_2$ calibration bottles from Air Liquide (France), containing 99.4 ± 2.0 ppm and 952 ± 18 ppb of CO. The gas from the bottles was mixed with pure $N_2$ at a flow rate of 100.0 ± 0.5 ml/min using a GB100 gas mixer (MCQ Instruments, Italy). The pressure was maintained at 100 ± 0.1 mbar with an MKS Instruments (United States) pressure controller. The temperature in the gas line was measured using a Thorlabs TSP01 sensor, and gas pressure was measured with a PAA-33X pressure sensor from Keller (Switzerland).

## Results and Discussion

This part of the paper discusses data processing for each technique and comparison of their performance. To determine the linear range of each technique, we conducted experiments with a 100 ppm CO bottle, as the transmission spectrum would be already saturated after ~65 ppm. This saturation allowed us to investigate whether dispersion measurements provide a wider linear range compared to absorption. Next, we characterised the performance of each method at lower concentrations with a 1 ppm CO bottle. From the calibration curves, we assessed the figures of merit. Finally, we evaluated the long-term stability of each method using the Allan-Werle deviation analysis.

Before each experiment, the mirror in front of the photodiode was realigned to reduce fringes. After alignment, the instrument was left idle for an hour to allow for the temperature to stabilise. All the spectra used for data processing are presented in Fig. 4(a), Fig. 5, Fig. 7, and in the Appendix Fig. A.1-2.

### Instrument Response

In this subsection, we discuss the performance of the techniques based on the raw spectra from the instrument. For TDLAS we looked at a transmittance, for WMS it was a 2*f* signal from the digital lock-in amplifier, and for HPSDS it was a 1*f* phase from the digital lock-in amplifier. In total, we measured spectra at more than 10 different concentrations ranging from 0 to 100 ppm.

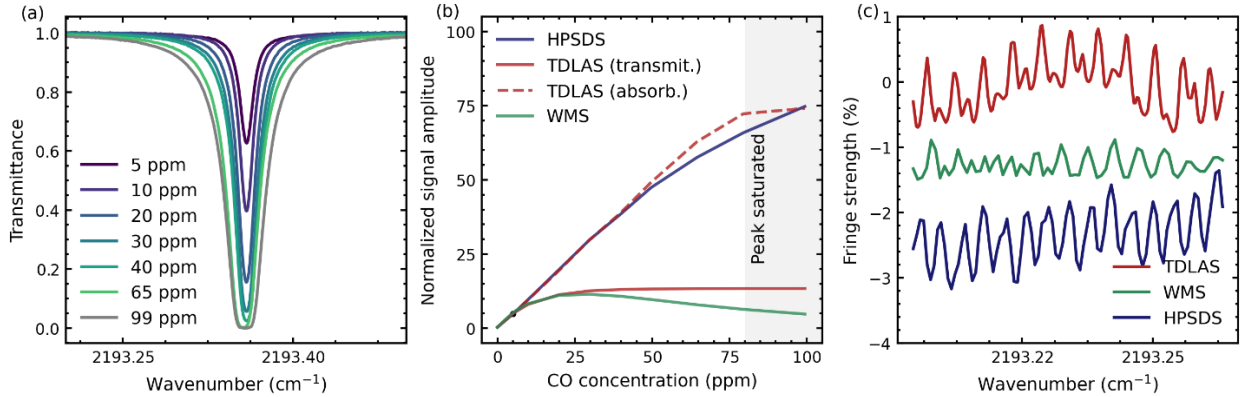

**Fig. 4.** (a) Measured transmittance demonstrated line saturation after 65 ppm. (b) Comparison of peak-to-peak amplitudes from 0 to 100 ppm for TDLAS (in transmission and absorption), WMS, and HPSDS. The observed behaviour in the WMS trend is due to the sample no longer being in the optically thin condition. (c) Fringe patterns in the non-absorbing region of the spectrum. The fringe strength for each technique is normalised to the observed signal amplitude and presented in per cent. All spectra were recorded at 0.75 ppm concentration. Each spectrum was averaged for 100 seconds.

First, we determined the concentration at which the transmittance saturates, which was at ~65 ppm, as seen in Fig. 4(a). Next, we plotted the peak-to-peak values of the measured spectral lines in Fig. 4(b). For visual comparison, the amplitudes were normalised at a 5 ppm value for all techniques. The TDLAS transmittance started to saturate after 50 ppm due to the exponential relationship between line peak and concentration. For WMS, the signal rose until 25 ppm and then dropped, displaying non-monotonic behaviour due to the sample no longer being in the optically thin condition. In contrast, the HPSDS phase monotonously rose up to 100 ppm.

After analysing the non-absorbing part of the spectra, we observed the fringes to be less than 0.04% in transmittance. In general, fringe amplitude was comparable for all the techniques, due to their optical nature, as presented in Fig. 4(c). From our experience, fringes for the HPSDS technique were the least sensitive to alignment, which could be attributed to the fact that HPSDS is known to be immune to power fluctuations.

## Basic Calibration Curves Based on Peak-to-Peak Evaluation

The key principle of laser spectroscopy is to transform a raw signal from a photodiode into concentration values. In a simplified scenario, the signal's peak-to-peak amplitude is recorded at different gas concentrations, providing a calibration relationship between spectrometer output and concentration, which is then used for field measurements. For calibration purposes, a standard gas bottle of 100 ppm carbon monoxide (CO) was dynamically diluted with dry nitrogen ($N_2$) in a stepwise manner using a mass flow controller. At each step, spectra were recorded for 1 minute at a 1Hz rate.

The resulting spectra for each technique are presented in Fig. 5. For TDLAS, we first removed the background $I_0(\tilde{v})$ using a conventional method by dividing the raw spectrum by the pure $N_2$ spectrum. We then converted the data to absorption by taking the natural logarithm. For WMS no data preparation was necessary. For HPSDS (dispersion technique) we subtracted the baseline, which was also measured with pure $N_2$.

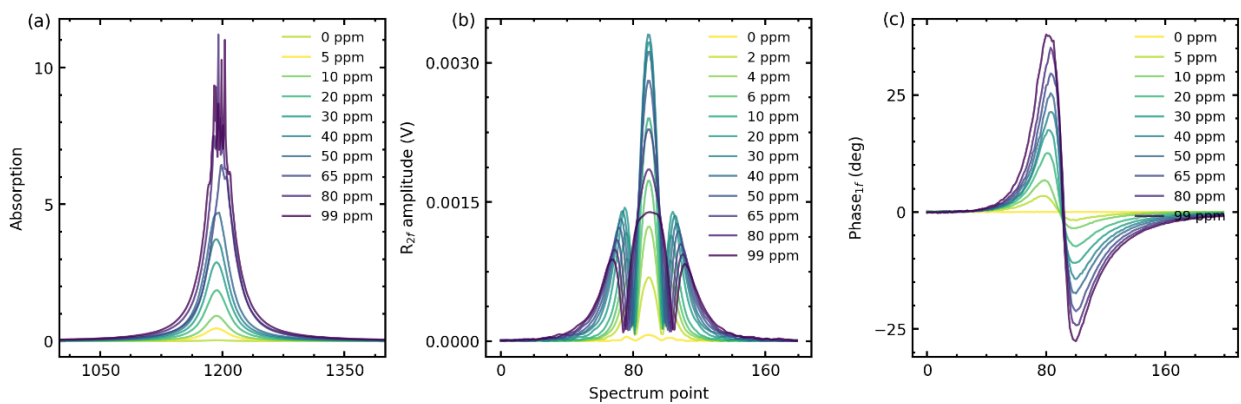

**Fig. 5.** (a) Absorption measurements using the TDLAS technique at various concentrations. Beyond 65 ppm, peak information was lost due to transmittance saturation. (b) 2*f* signal from a digital lock-in amplifier measured with the WMS technique. For concentrations above 20 ppm, the signal amplitude decreased while the linewidth increased. (c) The phase of the dispersion signal measured with the HPSDS technique. The phase amplitude increased monotonically without exhibiting saturation. We also note, that due to the imperfection of the gas mixer, we observed a weak leak in the CO mixer channel when a 100 ppm CO bottle was connected to the port, which resulted in ~0.2 ppm of CO even when measuring pure $N_2$. This leak did not affect further measurements of MDV and LOD, as we used a 1 ppm bottle, where the leak would be only ~2 ppb.

With the recorded spectra in Fig. 5, we calculated calibration curves based only on their peak values. 60 consecutive experimental measurements with a periodicity of 1 second were averaged for each concentration setpoint. The resulting linear regression of calibration curves is presented in Fig. 6 along with adjusted $R^2$. This experiment demonstrated that the WMS amplitude behaved linearly only under optically thin sample conditions up to 6 ppm. In contrast, HPSDS and TDLAS performed similarly, with TDLAS absorption showing the most linear response up to 65 ppm and HPSDS phase up to 50 ppm.

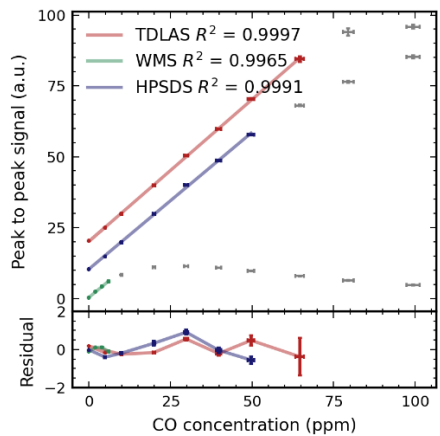

**Fig. 6.** Linearity assessment across a 100 ppm concentration range, including residuals. Only peak values were used in this analysis. For TDLAS, we calculated the amplitude of the absorption signal; for WMS, the amplitude of the *2f* signal; and for HPSDS, the amplitude of the 1*f* phase signal. Without applying any models, the TDLAS absorption signal demonstrated the linear range of 65 ppm, though limited by line saturation. The WMS technique, as expected, had the smallest linear range of 6 ppm, while HPSDS had a linear range of 50 ppm, similar to TDLAS.

## Spectral Models for the Different Techniques

As mentioned in previous subsections, in the *no-model* approach the amplitude of the spectra is recorded at different gas concentrations, providing a calibration curve, which is applied for field measurements. Unfortunately, the validity of such calibration is limited, as it depends on alignment, laser degradation, and gas parameters, such as temperature. In this subsection, we adapted a mathematical model of gas absorption and sensor response for each technique, which was then fitted to the measurements using non-linear least squares regression. For the gas model, we calculated the Voigt profiles with the HITRAN2020 database and the *hapi* Python library, with gas pressure and temperature included in the model [45,57]. This approach significantly improved the linear range and robustness of each method. The reader can find an in-depth explanation of the data processing steps in Appendix A.

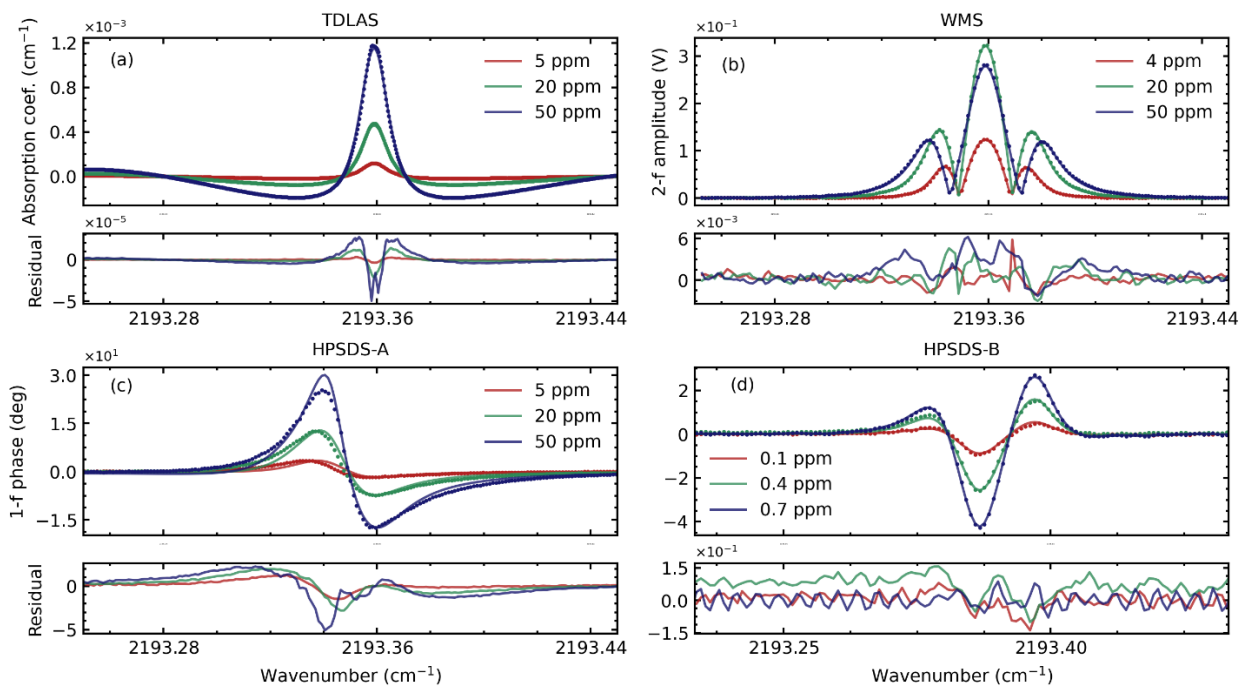

**Fig. 7.** Experimental measurements and spectral models for different spectroscopic techniques. Measured spectra are shown as dots and models as lines. The residuals are at the bottom. (a) For TDLAS, absorption signal was measured and fit with the simulation. Spectra look distorted due to baseline removal. (b) For WMS, a 2*f* signal was measured and simulated with the model. (c) For HPSDS-A, only the peak value of the measurement and spectral model was used. The laser was operated near the current threshold to extend the linear range. (d) For HPSDS-B, no spectral model was applied, instead we scaled the reference 0.75 ppm spectrum to fit other measurements. The laser was operated above the threshold current to improve signal-to-noise ratio at lower concentrations.

### Tunable Diode Laser Absorption Spectroscopy (TDLAS)

For TDLAS, we converted the raw spectrum into the absorption spectrum in a few steps, by first taking a natural logarithm, as in (1). Next, to retrieve the absorption spectrum, we applied Gram-Schmidt orthogonalization to generate a set of orthogonal polynomials that capture the low-frequency trends in the original spectrum, allowing for a reliable estimation of $I_o(\tilde{\nu})$. Traditionally, in transmission spectroscopy, the $I_o(\tilde{\nu})$ is estimated, and the spectrum is divided by it for normalisation, as in eq. (1). However, since we are working within absorption (after taking the logarithm of the non-normalised spectrum), this division translates to subtraction. Therefore, we subtracted the low-order polynomials from both the model and experimental data. By removing these low-order polynomials, we isolated the absorption features, effectively eliminating baseline $I_o(\tilde{v})$ contributions and power fluctuations, as $I_o(\tilde{v})$ is often defined using only a few polynomial orders.

$$\alpha_{baseline-free}(\lambda_i) = \alpha(\lambda_i) - \sum_{k=0}^{12} \alpha^k P^k(\lambda_i) \qquad (5)$$

The set of polynomials $P^k$ forms an orthogonal basis that meets the orthogonality condition. Detailed descriptions can be found in [20,21]. To eliminate DC noise, the laser current was swept with a 2 kHz ramp, which was averaged to 1 Hz spectra [51]. The wavenumber axis was established with Fabry-Perot etalon and measurements of air, where N₂O and CO lines served as absolute references, as in Fig. 2(b).

As shown in Fig. 7, a W-shaped residual was present, which is a common feature in a classic Voigt fitting [12]. This can be improved by using complex line shapes like the quadratic speed-dependent Voigt profile, but for that additional line profile measurements at different pressure levels would be required [12].

## Wavelength Modulation Spectroscopy (WMS)

Generally, WMS offers high precision but requires careful adjustment of laser parameters. In our study, we manually estimated the laser intensity and frequency modulation parameters. To model the recorded WMS spectra, we applied the following transformations to simulate the intensity on the photodiode $I_{PD}$ and the lock-in amplifier (LIA) output amplitude $R_{LIA}$:

$$I_{PD} = [I(t) + i_{mod}sin(2\pi\omega t)]\, e^{-\sigma[\tilde{v}(t)]NL} \qquad (6)$$

$$R_{LIA} = \sqrt{H(F(I_{PD}))^2 + F(I_{PD})^2}, \qquad (7)$$

where $i_{mod}$ (W) is modulation depth, $I(t)$ (W) is laser intensity, $\omega$ (Hz) is modulation frequency, $\tilde{v}(t)$ (cm$^{-1}$) is laser frequency tuning, $H$ is Hilbert transform, and $F$ is a bandpass filter function to isolate only the second harmonic of the signal (2*f*).

The laser current was swept at a rate of 1 Hz with an additional modulation at 10 kHz. This modulation was optimized to achieve the strongest response from the spectrometer. The photodiode signal was demodulated using a software-based lock-in amplifier in LabVIEW at the 2*f* frequency, and other LIA settings were configured to avoid smoothing of the line peak. The 2*f*/1*f* normalisation was out of the scope of this work.

## Heterodyne Phase Sensitive Dispersion Spectroscopy (HPSDS)

HPSDS signal represents the phase difference between three-tone laser radiation, caused by the anomalous dispersion of the gas. To simulate the phase, we first converted the Voigt line absorption spectrum to the refractive index with equation (4). Next, we applied a model (8) that calculates the phase response of the photodiode [35,36]. We manually estimated the following model parameters: strength of intensity modulation $m$, frequency modulation $\beta$, phase shift $\theta$ (deg) between them, and a final signal amplitude $K$ (deg). The output phase from the LIA is:

$$\phi = K * arctan \frac{2A\beta\, sin(\varphi_0-\varphi_1-\theta)-2B\beta\, sin(\varphi_{-1}-\varphi_0-\theta)+Am\, sin(\varphi_0-\varphi_1)+Bm\, sin(\varphi_{-1}-\varphi_0)}{2A\beta\, cos(\varphi_0-\varphi_1-\theta)-2B\beta\, cos(\varphi_{-1}-\varphi_0-\theta)+Am\, cos(\varphi_0-\varphi_1)+Bm\, cos(\varphi_{-1}-\varphi_0)}, \qquad (8)$$

where

$$\varphi_k = \frac{(\omega+k\Omega)L}{c}(n(\omega+k\Omega)-1) \qquad (9)$$

The $\varphi_0$, $\varphi_1$, and $\varphi_{-1}$ are the phases of the carrier wave and the two sidebands, $L$ is the path length, $\Omega$ is the modulation frequency. The coefficients $A$ and $B$ are equal to $exp[-\alpha(\omega+\Omega)L - \alpha(\omega)L]$ and $exp[-\alpha(\omega-\Omega)L - \alpha(\omega)L]$, respectively.

The laser current was swept at a rate of 1 Hz with additional RF modulation. As mentioned in the literature, the strongest system response could be reached when the modulation frequency is 0.58 of the full width at half maximum of the absorption line under investigation [58]. But in our case, the measured phase was influenced by laser IM-FM response, thus the modulation parameters were determined empirically. As discussed in the introduction, the laser's IM-FM behaviour can be controlled via the offset current. Thus we defined two modes of HPSDS operation, also illustrated in Fig. 7(c,d):

- **HPSDS-A [wide linear range]**: The laser operated close to its threshold current at *28.5°C* with RF modulation at 60 MHz and -6 dBm power (at 50 Ohm impedance). This mode provided the widest linear range but had poor SNR at low concentrations. Only peak values of the measurement and model were used since the model did not fit the entire shape of the experimental data optimally, as seen in Fig. 7(c).
- **HPSDS-B [best LOD]**: The laser operated far from its current threshold at *22°C* with RF modulation at 45 MHz and -3.3 dBm (at 50 Ohm impedance). In this mode, unwanted FM modulation amplified the signal peak, increasing the instrument response at lower concentrations, but decreasing the linear range. In this case, we replaced the HPSDS model with a scaling approach, where 0.75 ppm spectrum was used as a reference, and it was scaled to fit all other spectra.

Switching between these modes allowed us to either have a wide range of measurements, or high precision at low concentrations. The phase of the signal from the photodiode was demodulated using a lock-in amplifier in LabVIEW at a 1*f* frequency. The demodulation frequency was set at 10 kHz.

## Advanced Calibration Curves Based on Spectral Models

In this subsection we processed the same calibration measurements as before, but with the spectral models applied. At each concentration, spectra were recorded for 1 minute at a 1 Hz rate. In post-processing, each spectrum was fitted with a model described in the Spectral Models subsection, and then 60 consecutive experimental concentration values were averaged for each concentration setpoint. Finally, the observations in Fig. 8 were fitted with linear regression to present the residual and adjusted $R^2$.

In the case of a 100 ppm concentration range in Fig. 8(a), all three methods provided indistinguishable results in terms of linearity, with $R^2$ higher than 0.999. The linear range was determined by analysing the calibration curve responsivity threshold, as in ISO 8466-1-2021 Annex C. We note that the linear range for TDLAS was limited to 65 ppm, as the peak of the transmission line would saturate and produce errors when retrieving the concentration. We intentionally decided not to fit the transmittance directly, as it was more effective to remove the estimated background spectrum $I_o(\tilde{v})$ in absorption, after first taking the logarithm of the original spectrum, and then subtracting the background $I_o(\tilde{v})$, as described in the Spectral Models subsection. We also note that in the TDLAS regime, the measurements were verified and not calibrated, as we do not need prior knowledge of concentration to fit experimental spectra with the model of TDLAS. By virtue of this way of data analysis, TDLAS is often referred to as a calibration-free technique. As a result, TDLAS demonstrated a mean absolute error of 5.3% for the 100 ppm concentration range, which could be explained by the errors of the mixer (±0.5%), the reference bottle (±2%), and Voigt profile fit imperfection [12]. On the other hand, the spectral models for WMS and HPSDS were established by measuring standards, so they would output ppm values as seen in Fig. 8.

For the 1 ppm concentration range, all techniques again demonstrated similar linearity. We also note, that when we set the concentration of the gas mixer at 1 ppm, TDLAS was retrieving values of only 0.75 ppm, which could be explained by the interaction of CO with the bottle during long-time storage [59]. Thus, all other techniques were rescaled to 0.75 ppm for analysis in Fig. 8(b). We also mention that the laser was operating at different temperatures for HPSDS-B mode, as mentioned in the subsection on HPSDS model. The calibration curves in Fig. 8 were offset for clarity. Standard deviations of the method for both ranges are presented in Table 2.

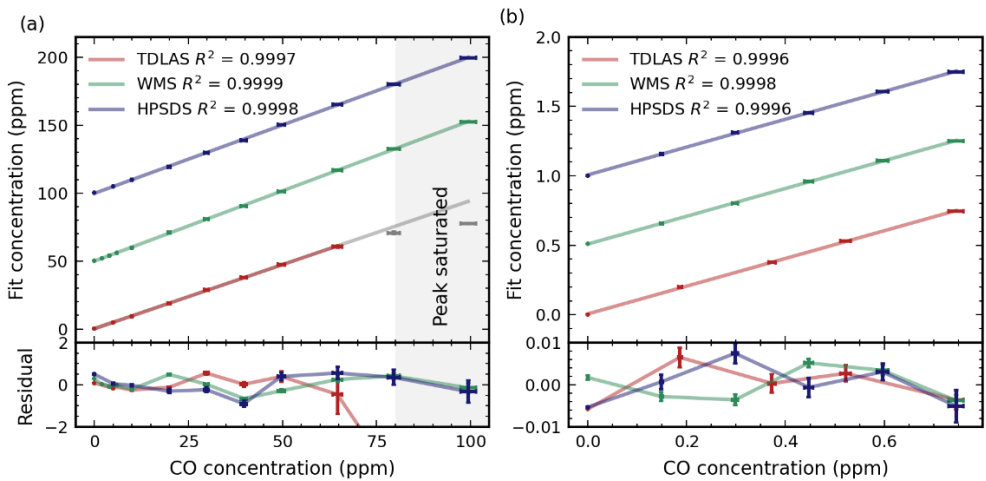

**Fig. 8.** The x-axis is the concentrations set at the gas mixer, the y-axis is the retrieved concentrations after spectral model fit. (a) Linearity assessment over the 100 ppm range. The TDLAS spectral model provided results without calibration, whereas WMS and HPSDS required calibration for the models to function. (b) Linearity assessment over the 1 ppm range and the residuals. The calibration curves for WMS and HPSDS were shifted for better visualisation in both plots.

## Long-term Stability

The stability of each technique was characterised by an Allan–Werle deviation analysis, as shown in Fig. 9 [56,60]. Measurements were processed with *AllanTools* Python library [61]. Assessments were based on 45 minutes of data recorded at 1-second resolution, taken at the end of each measurement interval. The stability experiments were conducted 3-5 times for each method. From the Allan-Werle plots, we determined the maximum averaging time still leading to improved sensitivity. The calculated values expressed in ppb correspond to 1 sigma of the calculated noise floor: TDLAS had a sensitivity of 140 parts-per-trillion (ppt) up to 100 seconds, WMS had a sensitivity of 250 ppt up to 80 seconds, and HPSDS had a sensitivity of 220 ppt up to 200 seconds. After these averaging times, drifts began to dominate the measurements leading to degrading performance in terms of achievable sensitivities, likely due to mechanical and optical instabilities of the instrument, as well as fluctuations in the mixer and pressure controller.

The narrow histogram for TDLAS could be attributed to power normalisation, which was omitted for WMS by design. The standard deviation of the HPSDS signal could have been influenced by additional RF components introducing noise. Our digital lock-in amplifier also could have contributed to extra noise. The software lock-in amplifier was chosen for the flexibility of switching between techniques, as the NI DAQ card was responsible for data acquisition. However, it could be concluded that the performance of the instrument was mostly limited by the fringe pattern, as suggested by the visual analysis of the noise in Fig. 4(c) demonstrating a periodic signal pattern attributed to the fringes.

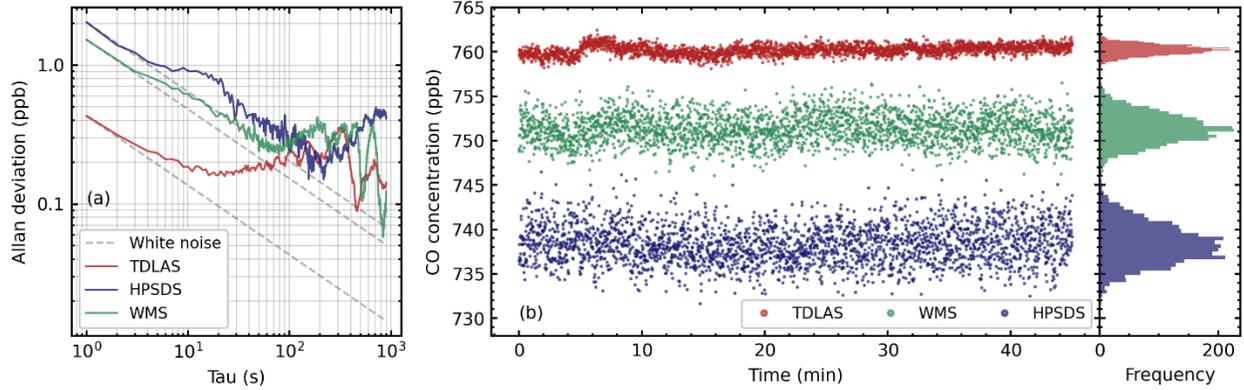

**Fig. 9.** (a) Allan–Werle deviation plots for all techniques at 0.75 ppm of CO. The white-noise behaviour (~$1/\sqrt{t}$, where $t$ is the integration time) is shown as a reference. (b) Time series of the last 45 minutes of measurements for the stability analysis, along with frequency of occurrence distributions. The measurements are shifted for visualisation purposes.

## Comparison

In this subsection, we present the comparison metrics as defined by ISO standards. The linear range was selected to characterise the measurement range, while the standard deviation and coefficient of variation of the method were used to assess the instrument's performance within that range. Additionally, the minimum detectable value and limit of detection were determined to evaluate the smallest measurable concentration. These calculations were based on the data shown in Fig. 8, with the results summarized in Table 2.

The linear range was determined using ISO 11843-2 and measurements in the 100 ppm range of concentrations. Such a procedure involves fitting the calibration data with a polynomial and identifying the threshold where the data points no longer follow a linear trend, as explained in Annex C of the standard.

The standard deviation of the method ($s_{x0}$) is a critical metric for evaluating the instrument's performance within the linear calibration range, expressed here in ppm. It is defined as $s_{x0} = s_y/b$, where $b$ (a.u./ppm) represents the slope of the calibration curve, and $s_y$ (a.u.) denotes the residual standard deviation of linear calibration which was calculated as follows:

$$s_y = \sqrt{\frac{\sum_{i=1}^{N}[y_i-(a+bx_i)]^2}{N-2}} \quad (10),$$

where $N$ is the number of calibration standard measurements, $a$ (a.u.) is the intercept of the calibration line, $y_i$ (a.u.) is the experimental data point. We note, that a.u. is arbitrary units.

We also define the coefficient of variation of the method as $V_{x0} = s_{x0}/\bar{x} \times 100$, where $\bar{x}$ (ppm) is the mean value of the standard concentrations $x_i$ (ppm). This metric is a relative measure of how small the standard deviation is compared to the measurement range. A smaller coefficient of variation indicates better instrument performance.

Finally, the minimum detectable value (MDV) was calculated using ISO 11843-2 with calibration data obtained from measurements in the 1 ppm concentration range [55]. In this context, the MDV represents the smallest value that can be reliably distinguished from zero with a specified probability. MDV reflects the sensitivity of the measurement process under defined calibration conditions. A similar popular metric of performance is a limit of detection (LOD), which is also defined in the Eurachem Guide as 3 times of standard deviation of concentration when measuring a blank or concentration close to the expected LOD [8]. In our case, LOD was calculated as 3 times of standard deviation of concentration measurements at 200 ppb for 60 concentration points during a minute.

**Table 2.** Comparison of techniques, related to the proposed realisation of techniques. The sensitivity from the Allan-Werle deviation was calculated via the *AllanTools* Python library for measurements with a 1 ppm bottle. The standard deviation of the method, linear range, and coefficient of variation of the method were calculated with ISO 8466-1. The minimum detectable value was calculated with ISO 11843-2 from the calibration data in the 1 ppm concentration range. The limit of detection was calculated according to Eurachem Guide from the measurements at 200 ppb. Pk-pk amplitude means peak to peak amplitude.

|  | TDLAS | WMS | HPSDS |
|---|---|---|---|
| **Linear range (pk-pk amplitude)** | 65 ppm | 6 ppm | 50 ppm (HPSDS-A) |
| **Linear range (spectral model output)** | 65 ppm | 100 ppm | 100 ppm (HPSDS-A) |
| **Standard deviation of the method (0…65 ppm)** | 0.44 ppm | 0.31 ppm | 0.58 ppm (HPSDS-A) |
| **Standard deviation of the method (0…1 ppm)** | 4.4 ppb | 4.0 ppb | 5.3 ppb (HPSDS-B) |
| **Coefficient of variation of the method (0…65 ppm)** | 1.6 | 1.4 | 2.1 |
| **Coefficient of variation of the method (0…1 ppm)** | 1.2 | 1.1 | 1.4 |
| **Minimum detectable value** | 15.2 ppb | 13.3 ppb | 17.6 ppb (HPSDS-B) |
| **Limit of detection ($3\sigma$ at 0.2 ppm)** | 3.2 ppb | 3.4 ppb | 4.4 ppb (HPSDS-B) |
| **Sensitivity (from Allan-Werle deviation)** | 0.17 ppb at 30s | 0.25 ppb at 70s | 0.20 ppb at 200s (HPSDS-B) |
| **Calibrated parameters** | None | Laser IM-FM response | Laser IM-FM response |
| **Advantages (with spectral model)** | - Straightforward data acquisition<br>- Calibration-free<br>- Intuitive signal shape<br>- Robust spectral model<br>- Good agreement with model | - Exceptional linearity<br>- Baseline free<br>- Good agreement with model<br>- No 2*f* saturation<br>- Possible measurements with multiple lasers[62] | - Wide linear range of raw signal<br>- Immunity to power fluctuation<br>- Less explored technique<br>- Less sensitive to fringes |
| **Disadvantages (with spectral model)** | - Saturates in transmittance<br>- Sensitive to signal distortion<br>- Sensitive to power fluctuation | - Dependent on laser parameters<br>- Requires calibration<br>- Sensitive to power fluctuation | - Requires RF components<br>- Strong dependency on laser IM-FM<br>- Complex mathematical model<br>- Requires calibration (if mixed IM-FM modulation is applied) |

# Conclusion

In this study, we demonstrated the capabilities of three different spectroscopic techniques for CO detection, based on absorption (TDLAS, WMS) and dispersion (HPSDS) spectroscopy. These techniques were integrated into a single experimental setup that allows switching between techniques with minimal modifications. By applying industry standard evaluation protocols our motivation was to explore and compare these techniques in search of a versatile approach capable of tackling multiple gas-sensing problems.

We initially compared the techniques based on the classical *no-model* approach, which relies solely on the peak amplitudes of the spectrum for instrument calibration. This comparison demonstrated that TDLAS and HPSDS surpassed WMS in terms of linear range. However, after applying the mathematical modelling of the spectra, all the techniques achieved comparable results in terms of limit of detection (LOD) ~0.2 ppb and linearity $R^2 > 0.999$, with long-term stability up to 100 s. Despite these similarities, each approach has its benefits and limitations. The fundamental limit of absorption techniques lies in signal saturation, while the dispersion approach relies on the coherence property of the laser source and measures the phase shift introduced by the sample's dispersion. The main advantage of the dispersion over absorption techniques is the extended linearity — demonstrated by the raw system response — and immunity of phase to power fluctuations [34].

Our analysis revealed that TDLAS provides straightforward data acquisition and a robust model, making it a reliable choice in unknown atmospheric environments. However, in our processing case, it exhibited a limited measurement range of up to 65 ppm caused by transmission peak saturation and was sensitive to fast power fluctuations. WMS demonstrated linearity up to 100 ppm when we applied the *2f* model. However, it was sensitive to incident power changes and required calibration of the model and simulation of the lock-in-amplifier signal. Its peak-to-peak detection approach could be beneficial for simplicity of data processing and baseline-free operation, when no line parameters are known (e.g. for benzene sensing). HPSDS, on the other hand, offered a wide linear range even without modelling the signal, and with the model, it reached a range of 100 ppm. We also demonstrated, for the first time, how after careful selection of the laser's regime of operation, HPSDS provides comparable precision to WMS and TDLAS. HPSDS stands out as a less explored technique, with potential in remote sensing, where a broad range of concentrations and particles in the air are expected. On the other hand, HPSDS does require a set of RF components and a complex mathematical model, which needs further research.

In conclusion, all three techniques are viable for CO detection, with the optimal choice depending on specific application requirements such as the measurement environment, expected concentration range, and instrument simplicity. Another factor is the background expertise of the researcher, as we encountered strong opinions when discussing this topic with individuals form the international scientific community of gas laser spectroscopy. Therefore, in our study, we established baseline performance metrics for each technique with industry standards, providing a foundation for informed selection in various gas-sensing applications.

# CRediT authorship contribution statement

**Iskander Gazizov:** Methodology, Software, Investigation, Validation, Writing - Original Draft, Visualization. **Davide Pinto:** Software, Investigation, Writing - Review & Editing. **Harald Moser:** Conceptualization, Resources, Writing - Review & Editing. **Savda Sam:** Investigation. **Pedro Martín-Mateos:** Methodology, Writing - Review & Editing. **Liam O'Faolain:** Project administration, Funding acquisition, Writing - Review & Editing. **Bernhard Lendl:** Conceptualization, Writing - Review & Editing, Project administration, Funding acquisition.

# Declaration of Competing Interest

The authors declare that they have no known competing financial interests or personal relationships that could have appeared to influence the work reported in this paper.

# Acknowledgements

This research was funded by the European Union's Horizon 2020 research and innovation program under the Marie Skłodowska-Curie grant number 860808, OPTAPHI.

# Appendix

TDLAS Data Processing:

1. Background Offset Removal: The background offset is subtracted by measuring intensity when the laser is turned off. The result is shown in Fig. A.1(a).
2. Wavenumber Calibration: The relative wavenumber axis is established by calibrating the spectrometer using a Fabry-Perot etalon. The absolute wavenumber is retrieved from peak positions listed in the HITRAN database.
3. Voigt Profile Simulation: Every 60 seconds, a new Voigt profile for CO and $N_2O$ is simulated based on the temperature and pressure reading inside the Herriot cell, as shown in Fig. A.1(b).
4. Baseline estimation: First, the natural logarithm of the spectrum from Step 1 is taken, followed by subtracting the baseline using orthogonal polynomials up to 12th-order (see equation (5)). This is equivalent to the calculation of $A(\tilde{v}) = \ln[I(\tilde{v})/I_0(\tilde{v})]$. The same procedure is applied to synthetic spectra from Step 3. Further details can be found in [20,21]. The absorption coefficient spectrum before and after baseline removal (in red and green, respectively) is shown in Fig. A.1(c).
5. Non-Linear Least Squares Fitting: The experimental spectrum (green) is fitted with a model spectrum (blue) using non-linear least squares regression to determine the concentration, as shown in Fig. A.1(c).

WMS Data Processing:

1. Spectrum Preparation: The raw 2*f* spectrum is cut and centered, as shown in Fig. A.1(d).
2. Wavenumber Calibration: The relative wavenumber axis scale is manually established during calibration, and the absolute wavenumber is retrieved from the HITRAN database.
3. Voigt Profile Simulation: Every 60 seconds, a new Voigt profile for CO is simulated based on the temperature and pressure reading inside the Herriot cell.
4. 2*f* Spectrum Simulation: the 2*f* synthetic spectrum is calculated using equations (6) and (7), with partial results shown in Fig. A.1(e). This 2*f* model requires calibration with prior knowledge of the CO concentration.
5. Concentration Fitting: The experimental spectrum is fitted using non-linear least squares regression to determine the concentration, as shown in Fig. A.1(f).

HPSDS Data Processing:

1. Phase Baseline Subtraction: The phase baseline is determined and subtracted based on $N_2$ blank measurements, as illustrated in Fig. A.1(g) and (j). This step is the same for both the HPSDS-A and HPSDS-B modes.

2. Phase Spectrum Preparation: The phase spectrum is cut and centered.
3. Wavenumber Calibration: The relative wavenumber axis scale is manually calibrated, and the absolute wavenumber values are retrieved from the HITRAN database.
4. Refractive Index Simulation: A Voigt profile is simulated based on the measurement conditions, and then converted to a refractive index spectrum using equation (4), as shown in Fig. A.1(h). This step is necessary only for HPSDS-A mode, thus there is a blank space for HPSDS-B in Fig. A.1.
5. HPSDS-A Fitting: The refractive index spectrum is converted to a phase signal using equations (8) and (9), then fitted with non-linear least squares regression to determine the concentration. The residual for the regression is defined as:

$$\varepsilon = |min(Exp) - min(Fit)| \quad (A.1)$$

Only the amplitudes of both spectra are used in concentration retrieval, shown in Fig. A.1(i).
6. HPSDS-B Fitting: A spectrum at 0.75 ppm CO concentration is selected as the reference and rescaled by multiplying it with a parameter $k$ to fit other spectra, as presented in Fig. A.1(k). The regression residual is defined as:

$$\varepsilon = |Exp - Ref * k| \quad (A.2)$$

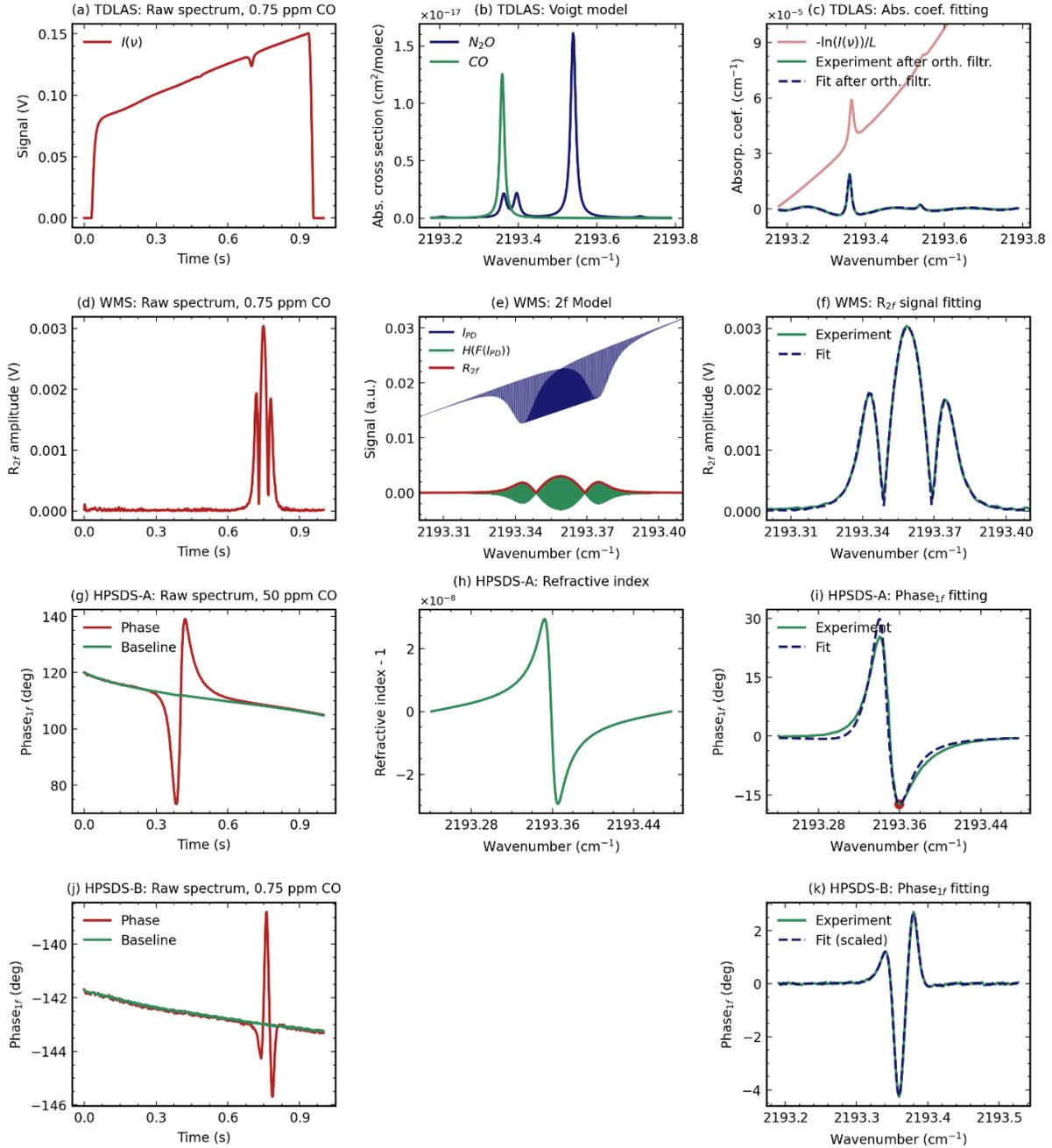

**Fig. A.1.** Overview of data processing steps for each spectroscopic technique. (a) TDLAS raw spectrum at 0.75 ppm of CO concentration in $N_2$ after averaging 2 kHz spectra into 1 Hz. (b) TDLAS Voigt profile simulation for CO and $N_2O$ based on temperature and pressure readings from the cell volume. (c) TDLAS absorption coefficient spectrum before (red) and after (green) baseline removal, with non-linear least squares fitting (blue) to determine concentration. (d) WMS raw $2f$ spectrum for 0.75 ppm of CO concentration. (e) Partial results of $2f$ synthetic spectrum simulation for WMS conditions, with a synthetic intensity of the photodiode signal (blue), partial results of $2f$ signal simulation (green), and final $2f$ amplitude (red). (f) WMS spectrum (green) fitted with $2f$ model (blue) to retrieve concentration. The output of this model requires calibration to function properly. (g) HPSDS-A phase (red) at 50 ppm CO concentration, and baseline (green) from the blank measurement. (h) The refractive index spectrum calculated from the absorption spectrum. (i) HPSDS-A spectrum fitting using only signal amplitudes for concentration retrieval. (j) HPSDS-B phase (red) at 0.75 ppm CO concentration, and baseline (green) from the blank measurement. (k) HPSDS-B phase spectrum fitting by scaling a 0.75 ppm reference spectrum.

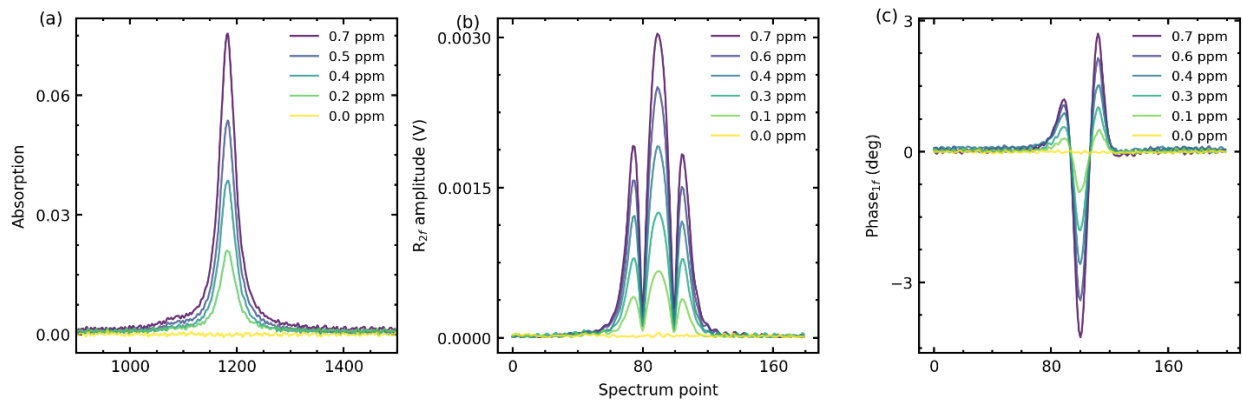

**Fig. A.2.** (a) Absorption measurements using the TDLAS technique in the concentration range of 1 ppm. (b) 2*f* signal from a digital lock-in amplifier measured with the WMS technique. (c) Phase of the dispersion signal measured with the HPSDS-B technique, when the laser operated far from the current threshold.